\documentclass[twocolumn, showpacs,preprintnumbers,amsmath,amssymb]{revtex4}

\usepackage{graphicx}

\begin{document}

\addtolength{\textheight}{1.2cm}
\addtolength{\topmargin}{-0.5cm}

\newcommand{\etal} {{\it et al.}}

\title{Towards the chemical tuning of entanglement in molecular nanomagnets}

\author{I. Siloi$^{1,2}$}
\author{F. Troiani$^2$}

\affiliation{$^{1}$ Dipartimento di Fisica, Universit\`a di Modena e Reggio Emilia, 
Italy}
\affiliation{$^{2}$ S3 Istituto Nanoscienze-CNR, Modena, Italy}

\date{\today}

\begin{abstract}

Antiferromagnetic spin rings represent prototypical realizations of highly correlated, 
low-dimensional systems. 
Here we theoretically show how the introduction of magnetic defects by controlled chemical substitutions results in a strong spatial modulation of spin-pair entanglement within each ring. Entanglement between local degrees of freedom (individual spins) and collective ones (total ring spins) are shown to coexist in exchange-coupled ring dimers, as can be deduced from general symmetry arguments.
We verify the persistence of these features at finite temperatures, and discuss them in terms of experimentally accessible observables. 

\end{abstract}


\maketitle

From a magnetic perspective, most molecular nanomagnets (MNs) can be essentially 
regarded as spin clusters with dominant exchange interaction \cite{gatteschi}. 
As such, they represent prototypical examples of correlated, low-dimensional quantum 
systems. It is thus tempting to consider how the wide tunability of their physical 
properties, enabled by chemical synthesis, can allow to control quantum entanglement 
\cite{amico,horodecki}.
In particular, different forms of entanglement can be possibly fine-tuned by chemical processing of well defined molecular building blocks, such as the controlled substitution of single magnetic ions or the growth of supramolecular bridges, 
both of which have been recently demonstrated in Cr-based wheels \cite{winpenny,timco}.
The former process results in the introduction of magnetic defects in otherwise 
homogeneous molecules \cite{larsen}, 
thus affecting correlations between their constituent spins.
The latter one can instead induce weak exchange couplings between two or more 
MNs \cite{vbellini}, so as to entangle their total spins \cite{candini10}. 

The effect of magnetic defects on entanglement has been 
widely investigated in infinite systems of 1/2 spins \cite{amico}.
With respect to these, MNs present some relevant differences, that make
them peculiar model systems: they typically consist of $s>1/2$ spins, are finite 
systems, and are
characterized by point symmetries  rather than translational invariance. 
The role played by these features \cite{troiani11}, as well as by the specific magnetic defects and supramolecular structures mentioned above, represents a specific reason of interest for the study of quantum entanglement in MNs.
Here we refer to a prototypical class of molecules, namely the heterometallic octahedral Cr$_7$M wheels, obtained by replacing one of the Cr ions in the parent Cr$_8$ molecule with different transition metals M \cite{caciuffo,bellini10}.
We theoretically show how the magnetic defects spatially modulate spin-pair entanglement, and identify a clear dependence of such effect on the impurity spin $s_M$. We finally investigate the 
interplay between intra- and inter-molecular entanglement in dimers of Cr$_7$M rings, whose compatibility can be deduced from general symmetry arguments. All these features are fully captured by local observables, such as exchange energy of individual spin pairs and partial spin sums, that are now available in direct geometry inelastic neutron scattering \cite{brukner,INS}.

%
\begin{figure}[ptb]
\begin{center}
\includegraphics[width=8.5cm]{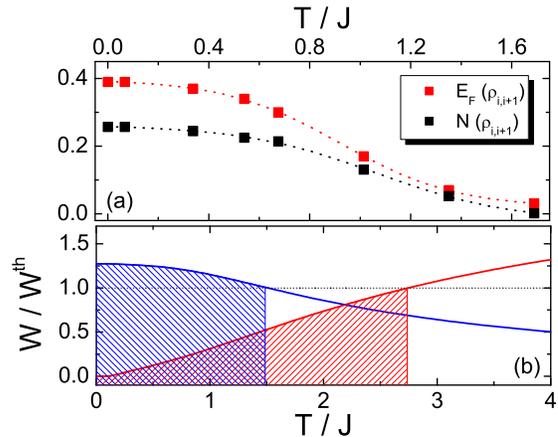}
\end{center}
\caption{(color online) 
(a) Negativity $\mathcal{N}$ and entanglement of formation $E_F$ for nearest neighboring spins of the Cr$_8$ ring, as a function of temperature. The dotted lines are the gaussian fits of the reported points.
(b) Temperature dependence of the entanglement witnesses $ W_S $ (red) and $ W_H $ (blue), normalized to the respective threshold values: $ W_S^{th} = 12 $ and $ W_H^{th} = - 18 J$. 
The shaded areas correspond to the temperature ranges where the two EWs detect entanglement.
}
\label{fig1}
\end{figure}
%

The dominant term in the spin Hamiltonian of the Cr$_7$M rings is the antiferromagnetic exchange between nearest neighbors \cite{troiani05}:
\begin{equation}
H = J \sum_{k=1}^{N-2} {\bf s}_k \cdot {\bf s}_{k+1} + 
    J'\,\, {\bf s}_N \cdot ( {\bf s}_{N-1} + {\bf s}_{1} ) ,
\label{ham}
\end{equation}
where $ s_{k < N} = s_{\rm Cr} = 3/2 $, $ s_N = s_{\rm M} $, and $N=8$. 
Additional terms, accounting for 
local crystal field and dipolar interactions, 
are typically two orders of magnitude smaller, and can be safely neglected in the present context. The spin $s_{\rm M}$ introduced by the 
chemical substitution determines the total spin $S$ of the ring ground state. 
In particular, the spin values for the elements M=Zn, Cu, Ni, Cr, Fe, Mn, are 
$s_{\rm M} = 0, 1/2 , 1 , 3/2, 2 , 5/2 $,
corresponding to 
$S   = 3/2 , 1 , 1/2 , 0 , 1/2, 1 $, respectively \cite{caciuffo,bellini10}. 

In order to investigate the entanglement properties of the Cr$_7$M molecules, we 
compute entanglement witnesses and measures \cite{horodecki,guhne09}.
An entanglement witness (EW) is an observable whose expectation value can exceed a given threshold only in the presence of (a particular form of) entanglement.
For a ring consisting of $N$ exchange-coupled spins $s$ (with $J>0$), the Hamiltonian itself can be regarded as an EW, being $ \langle H \rangle \ge -JNs^2 $  for any fully separable density 
matrix $\rho$ \cite{brukner04,toth05}.
The violation of such inequality implies the presence of entanglement within the 
system, and in particular between pairs of neighboring spins. 
We generalize the above criterion to the case of a ring with a spin 
$ s_N \neq s_{k<N} $ and two different exchange constants (Eq. \ref{ham}):
\begin{equation}\label{WU}
W_H \equiv \langle H \rangle 
\ge 
-(N-2) J s^2_{\rm Cr} - 2 J' s_{\rm Cr} s_{\rm M} 
\equiv W_H^{th} ,
\end{equation}
being $J,J' > 0 $.
The full separability of the density matrix also implies a lower bound for the 
variances of the total-spin projections \cite{wiesniak05}: 
$ \sum_{\alpha =x,y,z} [ \langle S_\alpha^2 \rangle - \langle S_\alpha \rangle^2 ] 
\ge Ns $.
In rotationally invariant spin Hamiltonians, the above quantity can be identified with the magnetic susceptibility, up to a multiplicative constant \cite{ghosh}. 
We modify the above criterion to account for the presence of the magnetic defect (Eq. \ref{ham}):
\begin{equation} \label{WS}
W_S \!\equiv\!\!\! \sum_{\alpha =x,y,z} [ \langle S_\alpha^2 \rangle - \langle S_\alpha \rangle^2 ]
\ge (N-1)s_{\rm Cr}+s_{\rm M} \equiv W^{th}_S .
\end{equation}
We note that, unlike $W_H$, $W_S$ depends on correlations between all spin pairs, and allows to detect forms of entanglement that don't show up in the two-spin density matrices.

While EWs represent a practical means for the detection of entanglement, its quantification requires entanglement measures \cite{horodecki}. 
In particular, entanglement between the spin pairs $ ( s_i , s_j ) $ is hereafter
quantified by the negativity $\mathcal{N}$ of their reduced density matrices
$\rho_{ij}$. This is defined as:
$ 
\mathcal{N} (\rho_{ij}) = (\sum_k |\lambda_k|-\lambda_k)/2 ,
$ 
where $ \lambda_k $ are the eigenvalues of the partially transposed density matrix 
$ \rho^{T_j}_{ij} $. 

%
\begin{figure}[ptb]
\begin{center}
\includegraphics[width=8.5cm]{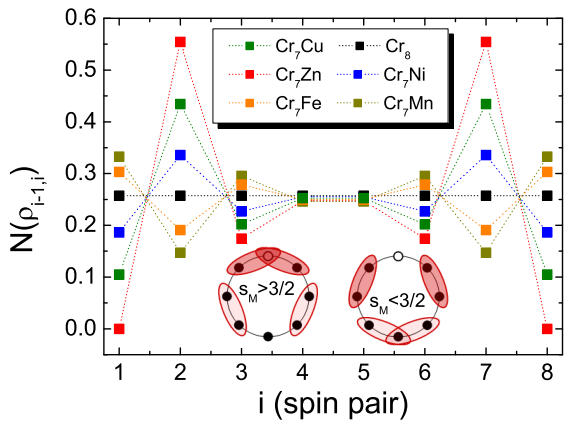}
\end{center}
\caption{(color online) Negativity of the neighboring spin pairs in the Cr$_7$M rings. 
The values refer to the ground state, with $M=S$, of the spin Hamiltonian $H$. 
The inset shows a pictorial representation of the entanglement localization induced 
by the defect M: the shaded areas highlight the most entangled spin pairs.}
\label{fig2}
\end{figure}
%

{\it Homometallic ring --- }
The parent molecule of the heterometallic Cr$_7$M rings is Cr$_8$ \cite{cr8}. 
Its energy spectrum is characterized by an $S=0$ ground state, separated from the first excited multiplet by a gap $ \Delta \simeq 0.559 J$ \cite{waldmann03}. 
As for the case of an $8$ qubit ring \cite{oconnor01,arnesen01}, spin-pair entanglement in the absence of an applied magnetic field is limited to nearest neighbors: 
we find in fact that $ \mathcal{N} = 0 $ at all temperatures for any other spin pair.
The temperature dependence of $ \mathcal{N} (\rho_{i,i+1}) $ presents instead a finite 
and nearly constant value for $ T / J \lesssim \Delta $, 
followed by a smooth decay at higher temperatures [Fig. \ref{fig1} (a), black squares]. 
The threshold value of the temperature at which $ \mathcal{N} $ vanishes is given by:
$ T_\mathcal{N} \simeq 1.58 J$. 
The same qualitative behavior is found for the entanglement of formation $E_F$ 
\cite{amico}, here reported as a benchmark (red squares).
The witness $ W_H $, like $ \mathcal{N} $, only reflects entanglement between neighboring spins, 
and in fact provides a similar value for the threshold temperature: 
$ W_H < W_H^{th} $ for $ T < T_H \simeq 1.5 J$ [Fig. \ref{fig1} (b)]. 
The witness $ W_S $ reveals instead further forms of entanglement, that persist up to higher temperatures, being $ W_S < W_S^{th} $ for $ T < T_S \simeq 2.75 J$. 
We finally note that values of $W_H/J$ lower than -20.65 and -21.8 imply the presence in the system state of three- and five-spin entanglement, respectively \cite{trosil}. 
In the Cr$_8$ ring, these thresholds are exceeded for $T/J$ lower than $1.05$ and $0.74$. 

{\it Heterometallic rings --- }
The chemical substitution of one Cr with an M ion introduces a magnetic defect in the molecule. This consists in the replacement of an $s_{\rm Cr}$ with an $ s_{\rm M} \neq 3/2 $ spin, and - in principle - in the modification of the exchange coupling 
between $s_8$ and its nearest neighbors (Eq. \ref{ham}). In the following we focus on the case $ J' = J $, which is compatible with the current estimates \cite{bellini10}, and allows to isolate the dependence on $s_M$ of spin-pair entanglement. 
In particular, we compute the negativity $ \mathcal{N} (\rho_{i-1,i})$ to quantify the ground-state entanglement, and the local witnesses $W_{H_i}$:
\begin{equation}\label{lews}
W_{H_i} \equiv 
J \langle {\bf s}_{i-1} \cdot {\bf s}_{i} \rangle
\ge -
J s_{i-1} s_{i} \equiv W^{th}_{H_i} .
\end{equation}
Violation of the above inequality implies entanglement specifically between $s_{i-1}$
and $s_i$. The witnesses $W_{H_i}$, that can be regarded as the local versions of 
$ W_H = \sum_{i=1}^8 W_{H_i} $, are used hereafter to estimate the threshold temperatures up to which entanglement persists in the various molecules.

The negativity corresponding to the ground state of $H$ is reported in Fig. \ref{fig2}. The translational invariance that characterizes the case of Cr$_8$ (black squares) is replaced by a strong spatial modulation. 
The negativity of the spin pairs with $i=1,3,6,8$ is an increasing function of $s_M$, while for $ i=2,7 $ the opposite occurs. For each molecule, the oscillations of $ \mathcal{N} (\rho_{i-1,i} )$ as a function of $i$ are paralleled
by those of $ W_{H_i} $ (not shown). In fact, they can be intuitively explained in terms of competition between the non-commuting exchange operators of $H$, since the state that maximizes entanglement between $s_i$ and $s_{i-1}$ coincides here with the one that minimizes their exchange energy \cite{entexc}. In particular, large values of the impurity spin tend to minimize $W_{H_1}$ (and thus to maximize entanglement of $s_{\rm M}$ with $s_1$), at the expense of the competing term $W_{H_2}$ (and of entanglement between $s_1$ and $s_2$). This in turn favors exchange interactions and quantum correlations between $s_2$ and $s_3$, and so on. The inverse applies to rings with $s_{\rm M} < s_{\rm Cr}$.

%
\begin{figure}[ptb]
\begin{center}
\includegraphics[width=8.5cm]{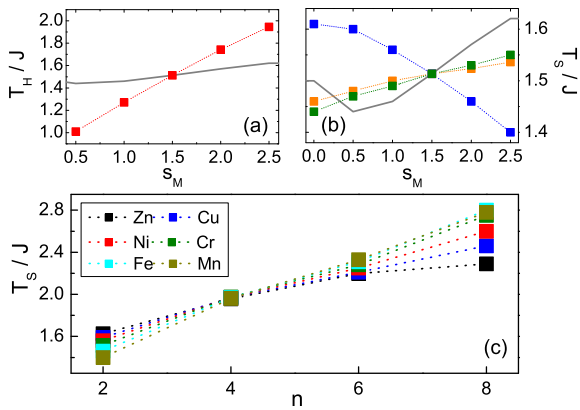}
\end{center}
\caption{(color online) Threshold temperatures of the local witnesses $ W_{H_i} $
as a function of the impurity spin $s_{\rm M}$: (a) M-Cr pair ($i=1$); (b) three inequivalent Cr-Cr pairs $i=2,3,4$ (blue, orange, green). 
The grey curve, reported in both panels, gives the threshold temperatures of the global witness $W_H$.
(c) Threshold temperatures for the witnesses $ W_{S_n} $ as a function of the 
spin number $n$ and of the impurity spin $s_{\rm M}$. 
}
\label{fig3}
\end{figure}
%

These general trends are confirmed by the finite-temperature results. In Fig. \ref{fig3} (a,b) we report the temperatures below which the witnesses $ W_{H_i} $ violate the inequalities Eq. \ref{lews}, thus detecting entanglement in the corresponding spin pair.
The spin pairs that exhibit the strongest dependence on $s_{\rm M}$ are the ones that include or are next to the magnetic defect: in particular,
$T_{H_1}$ increases linearly with $s_{\rm M}$ [panel (a), red squares], while $T_{H_2}$ decreases monotonically [panel (b), blue squares]. 
The other two inequivalent spin pairs also display a monotonic - though weaker - dependence on $s_{\rm M}$ (green and orange squares). 
These different behaviors cannot be appreciated in $ T_H (s_{\rm M}) $ (grey curve in both the upper panels). In fact, the threshold temperature of the global witness $W_H$ increases monotonically for finite values of the impurity spin, and systematically underestimates the persistence of entanglement, being $T_{H_i} > T_H$ for at least some $i$ in all molecules.

In order to bridge local (two-spin) and global witnesses, we introduce the EWs 
$W_{S_n}$, corresponding to the variances of partial spin sums:
\begin{equation}
W_{S_n} \equiv
\sum_{\alpha = x,y,z} [ \langle (S_{n,\alpha})^2 \rangle - \langle S_{n,\alpha} \rangle^2 ]
\ge \sum_{l=1}^{n} s_l \equiv W^{th}_{S_n},
\label{lWS}
\end{equation}
where 
$ {\bf S}_{n} = \sum_{l=1}^{n} {\bf s}_l $
and thus $W_{S_n} \equiv W_{S}$.
Their threshold temperatures $T_{S_n}$ are reported in Fig. \ref{fig3}(c) for even values of $n$. 
The most prominent feature is the significant increase of $T_{S_n}$ with $n$, for all the Cr$_7$M molecules. 
The same feature shows up within $n-$spin subensembles with odd $n$ (not shown).
Entanglement for 
$ T_H \lesssim T \lesssim T_S $
is thus progressively averaged away by the partial traces that are performed to derive the reduced density matrices 
$ \rho_n = {\rm Tr}_{n+1,\dots,8} \{ \rho \} $ 
of decreasing spin number $n$.
Based on calculations performed for the 8-qubit ring, we conjecture that
the density matrix $\rho (T)$ in such temperature range is given by the mixture of terms that include two-spin entanglement between different pairs. In other words, $\rho (T)$ would be a mixed 2-producible state \cite{guhne05}, whereas multi-spin entanglement is limited to $ T \lesssim J $ (see discussion on the Cr$_8$ molecule).
As to the dependence on the impurity spin, this is most significant for $n=2$ and $n=4$. In particular, $T_{S_2}$ decreases monotonically with $s_M$, like $T_{H_2}$ (that refers to the same spin pair), while $T_{S_8}=T_S$ displays the opposite behavior. 

{\it Ring dimers ---}
The Cr$_7$M rings are composite quantum systems, whose subsystems are represented by the constituents magnetic ions. However, they can also be regarded as building blocks of supramolecular assemblies, consisting of molecule dimers or oligomers \cite{timco}.
Here, the intermolecular exchange typically mixes states belonging to the ground $S$ multiplet of each ring, and can thus entangle collective degrees of freedom, such as the total spin projections of the MNs \cite{candini10}.

In the following, we investigate the interplay between individual- and collective-spin entanglement in a prototypical state of a Cr$_7$Ni-ring dimer:
\begin{equation}
|\Psi_{AB}^{Ni}(\alpha)\rangle =  \alpha |  1/2 , -1/2 \rangle - (1-\alpha^2)^{1/2} | -1/2 ,  1/2 \rangle ,
\end{equation}
where $A$ and $B$ label the two rings (with $S_{A/B}=1/2$), and the components $ |M_A,M_B\rangle $ are labelled after the values of the ring-spin projections ($S_{z}^A$ and $S_{z}^B$).
Entanglement between these collective spins can be quantified by the negativity: 
$ \mathcal{N}_{AB} \equiv \mathcal{N}(|\Psi_{AB}^{Ni}\rangle) = \alpha \sqrt{1-\alpha^2} $. 
The state $|\Psi_{AB}\rangle $ thus varies from a maximally entangled ($\mathcal{N}_{AB}=1/2$) to a factorizable state ($\mathcal{N}_{AB}=0$) as $ \alpha $ ranges from $1/\sqrt{2}$ to 1. 
%
\begin{figure}[ptb]
\begin{center}
\includegraphics[width=8.5cm]{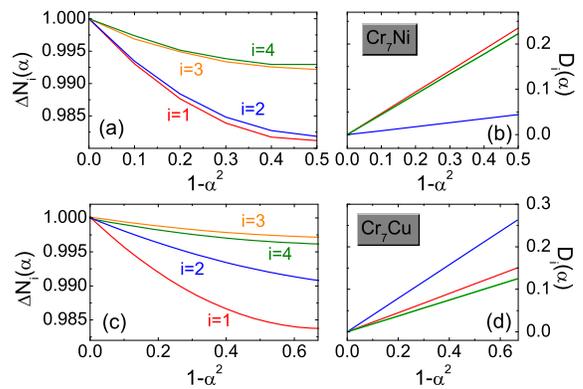}
\end{center}
\caption{(color online) 
Negativity of neighboring spin pairs belonging to Cr$_7$M ring dimers, with M=Ni (a)
and M=Cu (c): 
$ \Delta \mathcal{N}_i = 1 - \mathcal{N}_i (\alpha) / \mathcal{N}_i (0) $. 
(b,d)
Trace distance $D_i$ between the reduced density matrices $ \rho^A_{i-1,i} ( \alpha ) $ 
and $ \rho^A_{i-1,i} ( 0 ) $. The correspondence between spin pairs and colors is reported in panels (a,c). 
}
\label{fig4}
\end{figure}
%
Entanglement between two individual spins, $s_i^A$ and $s_j^A$, of ring $A$ is quantified by the negativity of the two-spin reduced density matrix $\rho_{ij}^A$: 
\begin{equation}\label{rhoijNi}
\rho_{ij}^A ( \alpha ) = \alpha^2 \rho_{ij}^{1/2} + (1-\alpha^2) \rho_{ij}^{-1/2} .
\end{equation}
Here, $ \rho_{ij}^{M_{A}} $ denotes the two-spin density matrix corresponding to the ground state of the single-ring Hamiltonian (Eq. \ref{ham}) with total-spin projection 
$M_{A}$.

Entanglement between the total spin projections of $A$ and $B$ ($\alpha < 1$) results in a mixing of $\rho_{ij}^{1/2}$ and $\rho_{ij}^{-1/2}$ (Eq. \ref{rhoijNi}). This generally tends to reduce entanglement between $s_i$ and $s_j$ - due to the convexity of entanglement measures \cite{guhne09} - making quantum correlation between individual and collective degrees of freedom mutually exclusive. As shown in Fig. \ref{fig4}, such reduction is however very limited in the case of exchange-coupled rings.
In fact, as $\alpha$ varies from $1/\sqrt{2}$ to $1$, the relative change of the negativity $\mathcal{N} (\rho_{i-1,i})$ is below 2\% for all the spin pairs (panel a),
in spite of the fact that the reduced density matrices $\rho_{i-1,i}$ vary
significantly with $\alpha$ (b). 
Such variation is quantified by the trace distance \cite{nielsen} between $\rho_{ij}(\alpha)$ and the reference state $\rho_{ij}(1)$:
\begin{equation}
D_{ij} (\alpha)\! =\! (1/2){\rm Tr} \sqrt{[\rho_{ij}(\alpha)\!-\!\rho_{ij}(1)]^\dagger [\rho_{ij}(\alpha)\!-\!\rho_{ij}(1)]} .
\end{equation}
Entanglement between individual spins within each Cr$_7$Ni ring ($s_i^A$ and $s_j^A$), is thus fully compatible with that between collective spins ($S_z^A$ and $S_z^B$), such as that induced by weak intermolecular exchange. 

The same applies to different ring dimers, such as that formed by two Cr$_7$Cu rings ($S_A=S_B=1$). 
Here, we consider the prototypical state 
\begin{equation}
|\Psi_{AB}^{Cu}(\alpha)\rangle = [(1-\alpha^2)/2]^{1/2} ( |1,-1\rangle + |-1,1\rangle ) - 
\alpha |0,0\rangle .
\end{equation}
This passes from maximally entangled ($\mathcal{N}_{AB}=1$) to factorized ($\mathcal{N}_{AB}=0$), as $\alpha$
varies from $1 / \sqrt{3}$ to 1, being
$ \mathcal{N}_{AB} = \alpha\sqrt{2(1-\alpha^2)}+(1-\alpha^2)/2 $. 
In this same range, the negativity displays a very limited decrease for all pairs of neighboring spins (panel c), also for significant values of the trace distance (d).

Symmetry arguments suggest that such compatibility between intra- and inter-molecular entanglement is more general. 
In fact, the entanglement witnesses $ W_{H_i} $ are scalar operators. 
As results from the Wigner-Eckart theorem \cite{tsukerblat}, their expectation value is thus identical for all the states that form an irreducible representation of the rotational group, such as the eigenstates belonging to the ground $S$ multiplet of the spin-ring Hamiltonian $H$ (Eq. \ref{ham}).
As a consequence, if any of the inequalities Eqs. \ref{WU},\ref{lews} is violated 
by a single-ring ground state, it's also violated by ring-dimer singlet states such as   
$ |\Psi^{Ni}_{AB}(1/\sqrt{2})\rangle $
and
$ |\Psi^{Cu}_{AB}(1/\sqrt{3})\rangle $,
where $S_z^A$ and $S_z^B$ are maximally entangled and the single-ring density matrix is a mixture of all the ground multiplet states.
The same argument applies to the witnesses $W_{S_n}$, provided that their state is rotationally invariant: in this case, $\langle S_{n,\alpha} \rangle = 0$, and $W_{S_n}$ can be identified with the expectation value of the scalar operator ${\bf S}_n^2$.

In conclusion, we have shown how the introduction of magnetic defects in the heterometallic Cr$_7$M rings introduces a strong spatial modulation of pairwise 
entanglement, that persists at finite temperatures. This suggests that suitably combined chemical substitutions can represent an effective means for engineering entanglement in molecular systems. Besides, we quantitatively show that in ring dimers entanglement between individual and collective spins can coexist, and deduce the generality of such property from symmetry arguments. The discussed features are fully captured by local observables acting as entanglement witnesses, such as the exchange energy  of spin pairs and the variances of partial spin sums, that are accessible to direct geometry inelastic neutron scattering. 

We thank M. Affronte and V. Bellini for useful discussions. We acknowledge financial 
support from PRIN of the Italian MIUR.

\end{document}